\documentstyle[11pt,newpasp,twoside]{article}
\def\edcomment#1{\iffalse\marginpar{\raggedright\sl#1\/}\else\relax\fi}
\reversemarginpar

\begin{document}
\title{When did the Large Elliptical Galaxies Form?}
\author{P. J. E. Peebles}
\affil{Joseph Henry Laboratories, Princeton University, Princeton
NJ 08544, USA}

\begin{abstract}
The simple reading of the evidence is that the large elliptical
galaxies existed at about the present star mass and
comoving number density at redshift $z=2$. This is subject to the
usual uncertainties of measurement and interpretation in
astronomy, but should be taken seriously because it is indicated
by quite a few lines of evidence. And it might be  
a guide to a more perfect theory of galaxy formation.   
\end{abstract}

\section{Issues}
Current ideas about structure formation suggest roughly half 
the large elliptical galaxies were assembled at redshifts less
than unity. The measurements do not rule this out, as noted 
in many observational papers, but a far simpler interpretation is 
that the large ellipticals formed well before $z=1$. 

The commonly discussed structure formation theory,
$\Lambda$CDM,\footnote{This assumes the mass of the universe is
dominated by cold dark matter, with gravitational growth of
structure out of primeval adiabatic, Gaussian, and near
scale-invariant density fluctuations in a low density
cosmologically flat universe. 
I adopt Hubble parameter $H_o=70$ km~s$^{-1}$ Mpc$^{-1}$, matter
density parameter $\Omega =0.25$, zero space curvature, and a 
cosmological constant that makes fractional contribution 
$\Omega _\Lambda = 0.75$ to $H_o{}^2$.} 
certainly deserves careful attention because of its dramatic
success in the interpretation of the  
temperature anisotropy of the 3K thermal cosmic background
radiation. But the test is limited, and aspects of the
theory seem problematic: small-scale structure (Moore et al. 1999
and references therein) and the void phenomenon (Peebles 2001). 
The case for early formation of the elliptical galaxies deserves
careful attention too, because it is the simple interpretation
of an impressively broad range of observations. 

The main competing pictures of galaxy formation, at high redshift
and low, often are termed monolithic and hierarchical. This makes
historical sense,\footnote{Thus the Partridge \&\ Peebles
(1967) fit of the Friedmann-Lemaitre cosmology to the Eggen,
Lynden-Bell \&\ Sandage (1962) picture for formation of the Milky
Way indicates galaxies formed at $z\sim 20$. 
Tinsley (1972) and Larson (1974) pioneered the modern picture for
the properties of an elliptical that formed as a galaxy  
of stars at high redshift. At another extreme, Kauffmann, 
Charlot, \&\ White (1996) discuss evidence that only about one
third of present-day early-type galaxies were assembled at $z=1$, 
roughly in line with what would be expected in $\Lambda$CDM.} but
can be confusing, because one can imagine  
galaxies form by hierarchical growth of structure either at high
redshift or low. I will take ``formation'' to be the
assembly of more than half the material now in the luminous parts
of a galaxy, gathered within a sphere of radius 30~kpc, let us
say. The epoch at which half the present-day stars formed may
precede or follow assembly. Either way, galaxy evolution may
continue well after formation as defined here. This includes
merging of satellite galaxies, as in the Ostriker-Tremaine (1975)  
``cannibalism'' picture, and more major mergers of galaxies that 
formed in tight pairs, as might be seen in clusters at high
redshift (van Dokkum et al. 2000) and low (Struble \&\ Rood
1981). It also includes new generations of stars from 
gas from the outer parts of the galaxy, or recycled from
evolved stars in the ``frosting'' picture of Trager et al. (2000).  

I limit this discussion to the formation of large elliptical
galaxies, or in some cases E plus S0 galaxies. There is a lot to be
learned from late-type galaxies, of course, but the limitation
helps focus the discussion on systems that present us 
with a fascinating pattern of regularities. The simplicity of the
early-type galaxies allows close exploration of rich details of
departures from the regularities. Both aspects, regularity and
complexity, may prove to be important guides to how and when the
galaxies formed and what that might teach us about the physics of
our expanding universe. 

Since I am more taken by the side of simplicity I should
acknowledge here that this certainly is not the whole story: the
departures from the patterns reflect significant differences
among the histories of galaxies (Toomre 1977; Schweizer 2000). We
see this in strongly disturbed late-type systems, whose remnants
likely will qualify as new early-type galaxies, in more
moderately disturbed ellipticals, that have gained mass by recent
accretion or mergers, and in Butcher-Oemler (1984) galaxies, that
show us morphological transformations. Aspects of this complexity
are reviewed in \S 2.2.    

The issue for the present discussion is whether the examples of
late formation show us the main way galaxies formed, or amount to
a perturbative effect on physical properties that were
established much earlier. Arguments of some years' standing for
the latter are that one sees large elliptical galaxies of old
stars at high redshifts (Oke 1971, 1984), and that subsequent
formation might be expected to erase the correlations of color
and heavy element abundance with luminosity (Faber 1973;
Visvanathan \&\ Sandage 1977; Ostriker 1980). The known  
patterns of regularities among early-type galaxies are much
richer now. I devote most of this contribution to a review of
the literature of these patterns, because I find them fascinating
and surely educational. Just what we might be learning is briefly
considered in \S 3.      

\section{The Observational Situation}

\subsection{Patterns at Modest Redshifts}
Bower, Lucey, \&\ Ellis (1992), and Terlevich, Caldwell, \&\
Bower (2001) discuss  the color-magnitude relation for
early-type galaxies in the Coma cluster: 
\begin{equation}
U-V = \hbox{constant} - 0.08 M_V\pm 0.05.
\end{equation}
I have taken the slope from Bower et al. and the rms scatter to be  
intermediate between that observed for E and for E$+$S0 galaxies.
The color-magnitude relation at redshift $0.5\la z\la 1$ in
clusters is analyzed by Ellis et al. (1997) and van Dokkum et al.
(2000; 2001b), and in the Hubble Deep Field, that might be
representative of ellipticals in 
groups, by Kodama, Bower, \&\ Bell (1999). These results 
offer three constraints on when the ellipticals formed: the 
scatter of colors is a measure of the scatter of mean ages of
the stars in different galaxies, and hence of the typical age; 
the evolution of the color with redshift is a measure of the
epoch of formation of the stars; and the scatter of colors
constrains the typical number of generations of random mergers.

The relation between the color and age of a population of old
stars is $d(U-V)/dt\simeq -0.025$ mag~Gyr$^{-1}$ (from Fig. 33
of Worthey 1994). The scatter in equation (1) thus translates to 
an rms scatter of mean star ages, $\sigma _t\sim 2$~Gyr, if
galaxies with the same luminosity have the same metallicity. It
is easy to imagine the scatter $\sigma _t$ in ages is larger than
the mean galaxy age, $t(z_f)$, as a result of galaxy formation
spread over several factors of two expansion, but difficult to
imagine the synchronization that would make $\sigma _t$
appreciably smaller than $t(z_f)$ (Bower et al. 1992). In the 
cosmologically flat model with the parameters adopted here 
the condition $\sigma _t\ga t(z_f)$ says $z_f\ga 3$. A similar
bound follows from the scatter of colors in cluster and group 
ellipticals at $0.5\la z\la 1$. This is an underestimate of the
bound on $z_f$ if the scatter in color has a contribution from
metallicity that is uncorrelated with age, an overestimate if
correlated fluctuations in age and metallicity cancel the effect
on color.    

The second measure is the variation of the zero point of the
color-magnitude relation with redshift. The measurements at $z<1$
match what would be expected if the redshift of formation of the
majority of field and cluster ellipticals satisfied $z_f\ga 2$ 
(Kodama et al. 1999). Van Dokkum \&\ Franx (2001) make the very
good point that this measure may be biased by morphological
transformations that remove the bluer galaxies from the accepted
sample of early types. Arguing against the bias is the
consistency with the limit from the scatter of colors. 
The bias also is tested by the 
evolution of the luminosity function: if a significant fraction
of present-day early-type galaxies entered the ranks at
$z<1$ then counts of early-type galaxies would show a deficit at
$z=1$. Kauffmann, Charlot, \&\ White (1996) find evidence for
this deficit. But in the deeper and perhaps more secure Groth HST
Strip, Im et al. (2000) find little evidence for 
evolution in the numbers of early-type galaxies back to $z=1$,
after allowance for the luminosity evolution expected for old
star populations. 

The third measure is less constraining but worth considering.
If three identical elliptical galaxies on the mean
color-magnitude relation merged to form an elliptical with three
times the luminosity (and the stars where the colors are
measured, at projected radius $\sim 4$~kpc in Bower et al. 1992,
fairly sampled the progenitors), it would produce an elliptical
that is two standard deviations off the mean relation for
galaxies that had avoided this round of merging. In the Kauffmann
\&\ Charlot (1998a) selective merging picture the color-magnitude
and other correlations can survive a considerable number of 
generations of mergers because large galaxies tend to merge with 
large galaxies and small with small, with limited scatter in
the numbers of generations of mergers. This is suggested by the
CDM theory for structure formation, and certainly could happen
at high redshift. But it is hard to see how merging could be
selective at the present epoch because maps of the nearby
galaxies show large and small are intermingled. Pending more
detailed observations of the situation at $z=1$, and analyses of
models for random merging, I would guess the tightness of the 
color-magnitude relation says fewer than half the ellipticals can
have undergone three major mergers since formation. 

Similar considerations apply to the fundamental plane and
luminosity-size relations. For early-type galaxies in clusters 
J\o rgensen, Franx, \&\ Kj\ae rgaard (1996) find
\begin{equation}
L\propto{\cal M} ^{0.76}r_e{}^{0.02} e^{\pm 0.23},
\end {equation}
where $L$ is the $r$-band luminosity, the mass is
${\cal M}\propto r_e\sigma ^2$, the half-light radius is $r_e$, the
velocity dispersion is $\sigma$, and the last factor in
equation~(2) is the rms scatter of individual galaxies from the
mean relation. This scatter is thought to be dominated by
intrinsic differences among galaxies. If the difference is a
scatter in ages  
then the relation between luminosity and age in an old star
population, $d\ln L/d\ln t \simeq 0.8$ (Tinsley 1972; Worthey
1994), translates to scatter $\sigma _t=4$~Gyr in ages. If 
$\sigma _t\ga t(z_f)$ this says $z_z\ga 1.8$,
comparable to the constraint from the evolution of zero point of
the color-magnitude relation. 

From the fit of the fundamental plane relation to cluster
galaxies at redshifts $z<0.6$, J\o rgensen et al.
(1999) find that the mass-to-light ratio varies with
redshift about as expected for the evolution of a star population
that formed at redshift $z_f\ga 2.5$. The fundamental plane
relation for ellipticals in groups is observed to 
be quite similar to that of cluster ellipticals (Kochanek et al.
2000; de la Rosa, de Carvalho, \&\ Zepf 2001; van Dokkum et al.
2001a), and yields a similar bound on the redshift of formation
of the stars.

The luminosities and radii of elliptical galaxies, in clusters
and the field, are correlated as 
\begin{equation}
L_{\rm B}\propto r^{1.33}.
\end{equation}
The evolution of the luminosity zero point with redshift again
agrees with the evolution of a star population that formed at
redshift well above unity (Barger et al. 1998; Ziegler et al.
1999; Schade et al. 1999).  

As for the color-magnitude relation, we can consider the
effect of merging on the fundamental plane and luminosity-size
relations. The constraint based on equation~(2) is not demanding
but worth recording. If three identical ellipticals with radii
$r_i$, all on the fundamental plane, merged to form an elliptical
with radius $r_f$, the 
resulting galaxy would be off the fundamental plane by the factor 
$L/L_{\rm FP}=1.30(r_i/r_f)^{0.02}$. This is only 1.3 standard
deviations, and not very sensitive to the ratio of initial to
final radii. If three identical galaxies on the size-luminosity
relation merged to form a galaxy on the relation, the final
radius would be $3^{0.75}$ times the initial radius. One can
imagine this is the natural value for the effect merging has on
radii, but one might imagine the kinetic energy of relative
motions of the galaxies is a random variable, that perturbs the
final radius and adds scatter to the luminosity-radius relation. 
Veeraraghavan \&\ White (1985) show that mergers of spirals
can produce a reasonable approximation to the elliptical
luminosity-velocity dispersion relation, and Hibbard \&\ Yun
(1999) show examples of galaxies in the late stages of mergers
that are likely to end up with the light profiles characteristic
of elliptical galaxies. Advances in the theory of radii of merger
remnants will be useful.

The merger rate is constrained also by the chemistry of large
ellipticals. This elegant application of astrophysics uses the
difference of time scales for production of the iron group
elements by thermonuclear burning in Type Ia supernovae, on the
order of 1~Gyr, and the considerably more rapid production of
$\alpha$-elements in massive  
Type~II supernovae (Thomas, Greggio, \&\ Bender 1999; Pagel 
2001). It is reckoned that about two thirds of the iron group
elements in the Sun came from Type Ia supernovae. This suggests
the material now in the thin disk of the Milky Way galaxy  
experienced 1~Gyr or more of star formation and element
enrichment before large-scale star formation in the thin
disk. In ellipticals, the abundance of magnesium and the
abundance ratio [Mg/Fe] are correlated with the velocity
dispersion, $\sigma$, and at large $\sigma$ is larger than solar
(J\o rgensen 1999; Colless et al. 1999). This 
is consistent with formation of large ellipticals on time scales
less than $1$~Gyr, as would happen if the ellipticals formed
at $z_f\ga 6$. The [Mg/Fe] relation is not so easy to reconcile
with the idea that ellipticals form by merging of spirals of
stars with systematically lower [Mg/Fe] (Thomas et al. 1999).  

Finally, in early-type galaxies and the bulges of spirals the
star velocity dispersion correlates with the black hole (or very
compact) mass in the nucleus (Ferrarese \&\ Merritt 2000;
Gebhardt et al. 2000). As we have noted, at $z<1$ large and small
galaxies are intermingled, so merging would broaden
the relation. That is avoided if merging is 
followed by significant black hole growth by accretion, at
a rate that might be controlled by the velocity dispersion.
But arguing against late growth is the existence of radio
galaxies and quasars at high redshift, that presumably operate
with black holes similar to those in radio galaxies at lower
redshifts.  

\subsection{Observations of Galaxy Formation at Low Redshift}
Interactions among galaxies, and of galaxies with extragalactic
gas and plasma, certainly have affected the population of
ellipticals at $z<1$; some aspects are reviewed here.  

Arp's (1966) Atlas of Peculiar Galaxies shows
examples of strongly disturbed late-type systems
that surely will merge. What else could that produce 
but elliptical galaxies (Toomre 1977; Schweizer 2000)? The nearby  
large elliptical galaxy NGC 5128, or Centaurus A, has a prominent  
disk of dust and gas, a clear example of a recent merger that
added about 10\%\ to the star mass (Israel 1998). 
Centaurus A may eventually merge with the other five spiral
members of the group, plus an S0 and some 20 dwarf and generally
gas-rich irregulars (C\^ ot\'e et al. 1997). That
seems likely to have a deleterious effect on the fit of 
Centaurus A to the patterns discussed in \S 2.1. Silva \&\ Bothun
(1998) show elegant examples of field ellipticals that may have
gained small central disks by merger-driven starbursts a few Gyr
ago, but they note that this applies to a minority of
ellipticals, and may qualify as frosting. I have not 
found an estimate of the fraction of large galaxies
at low redshift that show distinctive signatures of a substantial
gain of mass --- on the order of a factor of two --- from recent 
mergers. The commonly encountered list of prime candidates is not
long.   

There are young stars in ellipticals.
The fits of line strength indices to single-age star population
models yield a considerable scatter of ages of nearby early-type  
galaxies (J\o rgensen 1999; Trager et al. 2000). The indices are
sensitive to young stars, and show there has been significant
recent star formation. In the frosting model of Trager et al.
(2000) about 20\%\ of the stars in the central parts of nearby 
early-type galaxies are $\sim 5$~Gyr old, forming at $z\sim 0.5$,
the rest substantially
older. Menanteau, Abraham, \&\ Ellis (2001) find that the central
parts of about half the field ellipticals at 
redshifts $0.4\la z\la 0.8$ are relatively blue,
indicating recent star formation, again in the amount of a few
tens of percent of the star mass. The effect is not seen in
ellipticals in clusters, in line with the idea that field 
ellipticals are better able to retain gas shed from evolving
stars, that can make new generations of heavy element-rich stars.
Zepf (1997) made the excellent point that
if large ellipticals 
existed at $z>1$ in numbers comparable to today their colors
would have to have been affected by ongoing star formation,
in clusters and the field, because galaxies with the red colors
of a pure old population are rare at $z>1$. This ongoing star 
formation is a not unreasonable extrapolation of what is observed
at lower redshift, of course. I have not seen an observationally
based estimate of the redshift at which this ongoing star
formation might integrate to half the star mass in a typical
elliptical. 

Measurements of the fraction of S0 galaxies in clusters
show a distinct decrease from $z=1$ to the present (Dressler et
al. 1997; van Dokkum et al. 2001b). Following Dressler et al., 
one can imagine some cluster spirals, particularly those that
already have 
colors similar to early-type galaxies, transform morphology to
S0 as a result of disturbances by galaxies and the intracluster
medium, preserving the color-magnitude relation and modestly
disturbing counts of early-type galaxies. This is a clear
demonstration of the effect of environment on morphology, but not
necessarily an example of galaxy formation at $z<1$.  

\subsection{The Situation at Redshift Three}
This topic is left to last, as most speculative. But the striking
regularity of the $K$-band apparent magnitude-redshift relation
for radio galaxies beyond $z=3$ surely is telling something
of value about galaxy formation.  

The relation was discovered by Lilly \&\ Longair (1984) for 3CR 
radio galaxies. The colors of these objects at $z<0.5$ are
consistent with large present-day ellipticals, and at $z\sim 1$
the $r-K$ colors are consistent with Bruzual's (1981) models for
luminosity evolution with modest ongoing star formation. 
Van Breugel et al. (1998), Willott, Rawlings, \&\ Blundell
(2001), and Jarvis et al. (2001) show the $K$-band
magnitude-redshift relation extends beyond
$z=3$, without much increase in scatter at higher redshift.
At high redshift the optical and radio images tend to be
aligned, a sign of significant star formation or non-stellar
light associated with the radio activity. But the colors and
$K$-band magnitude-redshift relation 
to $z=3$ are consistent with the evolution of star populations 
that mostly formed at higher redshift, in giant galaxies of
stars. Consistent with this, a close examination of the spectra
of two radio galaxies at redshift $z\simeq 1.5$, by Nolan et al.
(2001), indicates ``star formation in at least these
particular elliptical galaxies was completed somewhere in the 
redshift range  $z = 3 - 5$.'' The Nolan et al. constraint from
spectra would allow the galaxies to be in pieces at $z=3$,
provided the subsequent assembly respected the patterns reviewed
in \S 2.1, but this is not consistent with the redshift-magnitude
relation.    

Since the star masses in high redshift radio galaxies seem to
be close to the largest present-day ellipticals,  
it is reasonable to assume these galaxies and their central black
holes generally have not grown much by merging or accretion since 
$z=3$. It seems reasonable also to suppose quasars at $z=6$
are associated with massive black holes, but more a matter of
speculation whether at $z=6$ the black holes are already clothed
with the present complements of stars at the present velocity
dispersion.

We have a measure of the numbers of radio-quiet ellipticals at
$z\sim 3$ from the deep rest-frame 
optical counts of Rudnick et al. (2001). They conclude that if
galaxies were conserved and their $B$-band luminosities a
factor of about three larger at $z=3$ than now it would about
match the observed counts. The factor of three would bring the
the $B$-band luminosity of the group elliptical Centaurus~A to 
$10^{10.7}L_\odot$, and the cluster elliptical M~87 to
$10^{11.3}L_\odot$. These are within the limit of the Rudnick et
al. survey. Thus the assumption that giant ellipticals are as
abundant at $z=3$ as now would be consistent with the Rudnick et
al. counts if the factor of three evolution of luminosity were 
astrophysically reasonable and the counts were not dominated by
something else. In the cosmology adopted here the ages of stars 
that formed at very high redshift are $t_3=2.3$~Gyr at $z=3$ and
$t_o=14$~Gyr now.  
Following Tinsley (1972), the evolution of luminosity is
$L_3/L_o\simeq (t_o/t_3)^{0.8}=4$. Worthey (1994) gives a similar
factor for $L_B$. Within the uncertainties, this seems consistent
with the Rudnick et al. (2001) factor of three. That is, by this
simple argument the counts are consistent with the
assumption that large ellipticals are conserved back to $z=3$. 
On the other hand, the passive evolution model of Kauffmann \&\
Charlot (1998b) predicts the galaxy counts at $z=3$ are well in 
excess of the observation. I await instruction on whether 
parameters might be tuned to fit the passive evolution model to
the Rudnick et al. counts, as the simple estimate of luminosity 
evolution suggests might be possible. 

The hypothesis of conserved ellipticals also is challenged by 
the star masses in Lyman-break galaxies at $z\sim 3$. Estimates
of star masses in the two nearby large ellipticals are
\begin{equation}
{\cal M}_{\rm star}({\rm Cen A})\simeq 10^{11.0}{\cal M}_\odot,
\qquad
{\cal M}_{\rm star}({\rm M 87})\simeq 10^{11.5}{\cal M}_\odot.
\end{equation}
The former is from Mathieu, Dejonghe, \&\ Hui (1996); the latter
assumes the same ratio of star mass to luminosity. Papovich,
Dickinson, \&\ Ferguson (2001) find the spectral
energy distributions of Lyman-break galaxies at $z\sim 3$ are
consistent with star masses in the range 
$10^9{\cal M}_\odot\la {\cal M}_{\rm star}\la 10^{11}{\cal
M}_\odot$. This reaches the mass of Centaurus A but not M~87. The 
straightforward interpretation is that the large ellipticals were
assembled as galaxies of stars at $z<3$. The evidence is that
radio elliptical galaxies formed earlier, however, and
Papovich et al. (2001) caution that estimates of the masses of 
old stars are subtle. Further exploration of this approach to an
upper bound on the redshift of assembly of ellipticals as large
as M~87 will be followed with interest. 

\section{So When did the Large Ellipticals Form?}
What does the $\Lambda$CDM theory predict? Application of
the theory on the scale of galaxies is difficult because the
behavior is complex. Here are three aspects of analyses of the
behavior. 

First, in the $\Lambda$CDM simulations
of Nagamine et al. (2001) the mean galaxy mass-to-light ratio in
the $B$-band has changed by about one magnitude from $z=1$ to the
present, close to pure passive evolution of an old star population.
Second, Thomas \&\ Kauffmann (1999) find that the abundance of
ellipticals at $z=2$ is predicted to be less than about one third
of what it is now. Third, Baugh et al. (1998) find that the
predicted comoving number density of galaxies with star mass 
$\ga 10^{10.1}{\cal M}_\odot$ has increased by a factor of about 50
since redshift $z=2$, and because evolution is expected to be more
rapid at larger mass this would say the number of elliptical
galaxies with masses in the range of equation~(4) is down by 
at least two orders of magnitude at $z=2$. 
The Nagamine et al. (2001) result is in quite satisfactory
agreement with the measurements of the fundamental plane as a
function of redshift; it shows us the tests for early 
formation must be considered with caution. The Baugh et al.
(1998) result suggests that if it were established that more 
than half the large ellipticals, at the range of star masses in 
equation~(4), are present at $z=2$, it would  
mean $\Lambda$CDM has to be adjusted. But impressions can change,
and we will be following developments. 

What do the observations say? The straightforward interpretation
of the measured variations with redshift of the color-magnitude, 
fundamental plane, size-magnitude, and $K$-band
magnitude-redshift relations is that the large  ellipticals in
groups and clusters had already formed by redshift $z=2$. This
relatively early formation is consistent with the 
straightforward interpretation of the 
tightness of the relations between color and magnitude,
[$\alpha$/Fe] and luminosity, and velocity dispersion and central 
black hole mass. As we have noted, Kauffmann-Charlot (1998a) 
correlated merging can keep these relations tight, but the
predicted segregation of large and small galaxies is not seen at
low redshift. Exploration of the situation at $z\sim 1$ will be
useful for analyses of the effects of mergers and to test
the $\Lambda$CDM prediction that large and small galaxies are
segregated at formation. 

Within the still substantial uncertainties of interpretation of
the observations and the theory $\Lambda$CDM is viable.
But it certainly seems to have problems with the epoch of galaxy
formation, as well as the formation of structure on  
smaller scales and the void phenomenon
on larger scales. One might also wonder about its application to
the cosmological tests, for it suggests a value of the mean mass
density that seems high compared to the measurement from weak
gravitational lensing (Wilson, Kaiser, \&\ Luppino 2001). We have
to straighten all this out before we can claim to have arrived at
a new era in cosmology, where we can be reasonably sure we know
what we're talking about.  
 
\acknowledgements
I have greatly benefitted by discussions with David Hogg,
Marijn Franx, and Masataka Fukugita, with whom I hope to write a
much more complete discussion of galaxy formation. This work is 
supported in part by the USA National Science Foundation.

\end{document}